\documentclass[preprint2]{aastex}
\received{2001 November 8}
\begin{document}

\def\arcdeg{\hbox{$^\circ$}}
\def\arcmin{\hbox{$^\prime$}}
\def\arcsec{\hbox{$^{\prime\prime}$}}
\def\lesssim{\mathrel{\hbox{\rlap{\hbox{\lower4pt\hbox{$\sim$}}}\hbox{$<$}}}}
\def\gtrsim{\mathrel{\hbox{\rlap{\hbox{\lower4pt\hbox{$\sim$}}}\hbox{$>$}}}}

\title{The Canada-UK Deep Submillimeter Survey V: The Submillimeter Properties of Lyman-Break 
Galaxies}

\author{T.M. Webb\altaffilmark{1}, S. Eales\altaffilmark{2}, S. Foucaud\altaffilmark{3}, S.J. 
Lilly\altaffilmark{4}, H. McCracken\altaffilmark{3}, K. 
Adelberger\altaffilmark{6}, C. 
Steidel\altaffilmark{5}, A. Shapley\altaffilmark{5}, D.L.
Clements\altaffilmark{2}, L. Dunne\altaffilmark{2}, O. Le F\`evre\altaffilmark{3},  M. 
Brodwin\altaffilmark{1},  
W.
Gear\altaffilmark{2}}

\altaffiltext{1}{Department of Astronomy and Astrophysics, University of Toronto, 60 St George St, 
Toronto, 
Ontario, Canada, M5S 1A1}
\altaffiltext{2}{Department of Physics and Astronomy, Cardiff University, P.O. Box 913, Cardiff, CF2 
3YB, UK} 
\altaffiltext{3}{Laboratoire d'Astrophysique de Marseille, Traverese de Siphon, BP8, 13376 Marseille 
Cedex 
12, France}
\altaffiltext{4}{Institut f\"{u}r Astronomie, ETH H\"{o}nggerberg, HPF G4.1, CH-8093, Z\"{u}rich, 
Switzerland}
\altaffiltext{5}{Palomar Observatory, California Institute of Technology, MS 105-24, Pasadena, 
California, 
USA, 91125}
\altaffiltext{6} {Harvard-Smithsonian Center for Astrophysics, 60 Garden Street, Cambridge, MA, 
USA, 
02138}

\begin{abstract}
We have used 850$\mu$m  maps obtained as part of  the Canada-UK Deep Submillimeter Survey 
(CUDSS) to investigate the sub-mm properties 
of  Lyman-break galaxies (LBGs).  We used three samples of Lyman-break galaxies: two  from the 
Canada-France Deep Fields (CFDF) survey 
covering CUDSS-14 and CUDSS-3, and one from Steidel and collaborators also covering  
CUDSS-14. 
  We measure  a 
mean flux from both CFDF LBG samples at a level of $\sim$2$\sigma$ of 0.414 $\pm$ 0.263  mJy for 
CUDSS-03 and  0.382 $\pm$ 0.206 mJy 
for CUDSS-14, but the Steidel et al. sample is consistent with zero flux.  From this we place upper 
limits 
on the Lyman-break contribution to the 
$850{\mu}m$ background of  $\sim$20\%. We have also measured the cross-clustering 
between the LBGs and SCUBA sources. From this measurement we infer a large clustering amplitude 
of 
$r_o$ = 11.5 $\pm$ 3.0 $\pm$ 3.0 
$h^{-1}$Mpc for the Steidel et al 
sample (where the first error is statistical and the second systematic), $r_o$ = 4.5 $\pm$ 7.0 $\pm$ 
5.0 $h^{-1}$Mpc for CFDF-14 and $r_o$ = 
7.5 $\pm$ 7.0 
$\pm$ 5.0  $h^{-1}$Mpc for CFDF-3. The Steidel et al sample, for which we have most only 
significant 
detection of 
clustering is also the largest of the three samples and has spectroscopically  confirmed redshifts  
\end{abstract}


\section{Introduction}
Recent work at optical and submillimeter (sub-mm) wavelengths has granted us unprecedented access 
to 
the high-redshift universe.  In particular,
the Lyman-break selection technique \citep{ste93} has provided thousands of star forming galaxies at  
redshifts z$\sim$3 and 
z$\sim$4.  These data, and the results of other optical surveys \citep{lil96,mad98,hog98}, have shown 
that 
the global star formation rate  (inferred  
from the $UV$ luminosity  density at different redshifts),  increases with redshift 
to  z$\sim$1 and may remain constant to at least z$\sim$4 \citep{saw97,ste99}, implying the beginning 
of the 
epoch of galaxy 
formation occurred at $z>>$ 1.  \

Studies of the spatial distribution of Lyman-break galaxies  show that they  are 
highly clustered even at these early redshifts \citep{gia98}.  This was initially unexpected since the 
clustering 
of galaxies had been shown to 
decrease 
with redshift to z$\sim$1 \citep{carl97, lefev96} in line with theoretical predictions, but the strong 
clustering of LBGs is  actually a natural 
consequence of the effects of bias. \citet{kai84} showed that the high 
peaks of the density distribution in the  early universe will have been highly clustered, and so objects 
that 
form from these high peaks, clusters at a 
redshift of zero or galaxies at a  redshift  of 3, should also be highly clustered.  \

The high star formation rates (50-100 M$_{\odot}$ yr$^{-1}$) and comoving density of the 
Lyman-break galaxies make them attractive  
progenitors of 
present-day elliptical 
galaxies \citep{pet98} though their masses are still highly uncertain \citep{saw98,som01}. However, the 
Submillimeter Common-User Bolometer Array (SCUBA) on the 
James Clerk Maxwell Telescope 
(JCMT) has revealed a population of dusty galaxies with implied star formation rates of $>$ 300 
M$_{\odot}$ yr$^{-1}$ 
\citep{sma97,hug98,eal99}.  The 
redshifts of these  objects are still  highly uncertain, with estimates of the median redshift  of the 
population lying between 2 and 3 
\citep{eal00,bar99,sma00,yun02,fox02}.  They have  similar spectral energy distributions  to  today's 
ultraluminous infrared  galaxies (ULIRGs) and often show 
disturbed morphology or multiple  components,  implying they may be the result of galaxy mergers 
\citep{lil99,ivi00}.   Both LBGs and SCUBA 
sources have sufficient star formation rates to form  present-day elliptical galaxies but the extremely 
high 
star formation rates of the latter mean 
this can be done on the order of 10$^8$ years, as the homogeneous properties  of local ellipticals 
indicate 
is the case.  \

The nature of the relationship between these two populations remains unclear.
An obvious scenario is one in which they form a continuum of objects with the bright sub-mm selected 
sources representing the highest star 
forming Lyman-break galaxies.   \citet{adel00} claim that by assuming an  
$L_{bol,dust}/L_{UV}$ typical of normal starbursts, the 
LBG population can produce the  bulk  of the  850 $\mu$m background. In this picture the two  
populations are the same objects and a separate 
population of  highly obscured ULIRG-like objects is not needed.    The ratio of optical 
to sub-mm emission for the brighter SCUBA sources is, however, much less than for the starbursts 
considered by Adelberger \& Steidel and so 
these objects almost certainly represent a separate population \citep{gea00, eal00,dow99}, but it is 
unclear 
whether the fainter SCUBA sources, 
with S$_{850} <$ 3 mJy, overlap with the LBG population. \

Various optical techniques have been used to infer the 
dust content of LBG's. \citet{pet01} have used optical line ratios  and \citet{sha01} fitted 
the 
predictions of star-formation models to  optical and near-IR photometry. Both have concluded that 
the 
most intrinsically luminous LBGs,  
which have higher star formation rates, contain more dust.  However,  a more reliable way of 
measuring 
the dust content is directly through 
sub-mm photometry. \citet{cha00} have used SCUBA to observe a sample of  high-SFR 
LBGs 
and they estimate that the 850$\mu$m 
flux density is at least two times lower than predicted from $UV$ colours. Using the sub-mm map of 
the 
Hubble 
Deep Field (HDF), \citet{pea00} statistically detected the sub-mm flux of galaxies with high $UV$ 
luminosities, and thus high star 
formation rates.  They detect a higher mean flux of 0.2 $\pm$0.04 mJy  (for galaxies with an inferred  
star 
formation rate (SFR) of 1 
$h^{-2}$M$_{\odot}$ yr$^{-1}$).   \

 This paper examines  the relationship between Lyman-break galaxies and SCUBA sources in  two 
Canada-UK Deep Submillimeter Survey 
(CUDSS) fields and is organised as follows.  \S 2 describes the sub-mm and optical/$UV$ data.  In \S 
3 
we investigate the sub-mm fluxes 
of  LBGs.   \S 4 discusses the dust properties of  LBGs that can be inferred from the results in \S 3.    
In 
\S 5 the correlation function 
between the two populations is presented.  In  \S 6 we discuss the results and their implications.\

\section{The Data}

We have mapped two areas within the  Canada-France Redshift Survey fields (CFRS), CFRS03+00 
and 
CFRS14+52, using 
SCUBA at 850$\mu$m.  This study was designed to be a blank-field survey for sub-mm-selected 
sources 
above 3 mJy but also 
produces  statistical information  for  objects below this flux.  Each map is roughly  6$\times$8  
arcmin$^2$ 
 and is a mosaic of  single jiggle maps.  The sub-mm data and the goals and results of the CUDSS  are 
discussed in detail in 
\citet{eal99,eal00,lil99,gea00,web02}; Webb et al, in preparation; Clements et al., in 
preparation. \

Surveys for Lyman-break galaxies have been performed over both these areas.   In the 14-hour field 
(CFRS14+52) we have access to Lyman-break data from 
two  separate sources, the Canada-France Deep Field survey (CFDF)  (\citet{mcc01}; 
Foucaud et 
al., in preparation)  and 
Charles Steidel and collaborators (Steidel, C. 2001, private communication).  The Steidel et al list 
contains 
86 galaxies within the SCUBA map area 
and most of 
these have spectroscopically confirmed redshifts.   It is an $\mathcal{R}$-limited survey with 
$\mathcal{R} 
< $ 25.5 (see \citet{ste93} for discussion 
of the  filter system) and is roughly 0.5 magnitudes deeper than the CFDF data. \

 The CFDF survey, and in particular the construction of catalogues and limiting magnitudes, is 
described in detail in \citet{mcc01}.  Details of the CFDF Lyman-break selection technique are found in 
Foucaud et al. (in preparation).   The 
survey is  $I$-limited  
 with  $I_{AB} < $ 24.5 and has selected galaxies over the redshift 
range 2.9 $< z < $ 3.5.  An  additional constraint of $(V-I)_{AB}<$ 1.0 has been introduced which  
reduces contamination by stars and elliptical 
galaxies at 
$z{\sim}$1.5. The CFDF team estimate (from simulations) the 
contamination due to stars to be at the 5\% level and that due to galaxies below $z{\sim}$2.9 to be 
15\%.  
The total level of contamination of 20\% 
is comparable to that of the Steidel sample.  In the redshift range of 2.9 $< z <$ 3.5 and with 
$I_{AB}<$ 
24.5 the CFDF method 
recovers 70\% of the Steidel et al catalogue.  Of the entire CFDF survey a 
subset  of 26 galaxies fall within our 14-hour SCUBA area and 29 within the  3-hour field area. \

\section{Submillimeter Flux of Lyman Break Galaxies}

\subsection{Statistical Results}

The CUDSS's large, contiguous maps provide a unique opportunity to statistically study the sub-mm 
flux 
of a relatively large number of 
Lyman-break galaxies.   Our method is simple:  we measure the sub-mm flux at each LBG position in 
the 
SCUBA maps and take a weighted mean 
of these data, obtaining the weights from the noise maps for each field (see \citet{eal00,web02}).   For 
high 
signal-to-noise measurements one 
would measure the flux at 
the peak of the beam (or the point-spread-function) but as these objects are all below the noise level 
we 
cannot locate the peak and must use the 
value at the LBG location (which will be, on average, offset from the peak as discussed below).  As we 
will 
show, it is quite crucial to first remove all 
sources  from the maps that are definitely not associated with an LBG.  \

Simulations in an earlier paper \citep{eal00} of the sub-mm data  have shown that the offset in our 
maps 
between the actual position of an object 
and its recovered position (for the $S_{850{\mu}m} > $ 3 mJy)  is usually within 6 arcseconds with the 
peak of the distribution occurring at 
approximately 3 arcseconds (also see \citet{hog01} for discussion of offsets).  The recovered position 
lies 
further than 8 arcseconds in only 5-10\% of cases, and so we removed all SCUBA 
sources  from the maps for which there was  no LBG within 8 arcseconds.  \

For the 3-hour field we removed all of the SCUBA sources, for the Steidel 14-hour sample, all but 
three 
sources and for the CFDF-14,  all but one source.  These four sources are listed in Table \ref{tbl-1} 
along 
with their identification probability.   The source removal, 
or {\it cleaning}, discussed in \citet{eal00}, includes the removal of the entire beam template.  That is, not 
only is the positive flux from a source 
removed 
but so is the negative flux due to the chop.  This step is vital since the beam profile, convolved with a 
point source, essentially extents over $> 75$ 
arcseconds in RA and can therefore interfere in flux measurements of nearby sources if not removed. \

We determined the probabilities that these four associations are simply chance coincidences in the 
following way.   If an LBG lies {\it d} arcsec 
from a SCUBA source, the  probability that it is unrelated to the SCUBA source is 
$P=1-exp(-{\pi}nd^2)$, where $n$ is the surface 
density of LBGs (the  more sophisticated analysis which takes account of the the magnitudes of 
galaxies is 
described in Webb et al. (in preparation). The 
probabilities, which are given in Table 2, are not particularly low and so the associations may not be 
genuine; but they {\it may}  be genuine and 
so we cannot remove these sources from the maps.  \

We can use the fact that  four LBGs do lie within 8 arcseconds of a SCUBA source to 
estimate how many real SCUBA-LBG associations we may have missed by discarding all SCUBA 
sources 
at a greater distance from an LBG. 
On the assumption that these four are genuine associations, we would expect, from the results of the 
simulations, that 0.05-0.1$\times$ 4 
associations to have an offset of $>$ 8 arcseconds.  Since some of these four associations are 
undoubtedly due to random chance, this gives an 
upper limit of 0.2-0.4 missed identifications. \

After removing all the sources except the above, we measured the noise-weighted mean of the flux at 
the 
positions of LBGs, treating these three 
samples of LBGs separately.  The results are summarised in Table \ref{tbl-2}.  For each sample, the 
mean 
flux is above zero, but clearly not at a very 
convincing level.  The significance of these measurements can be estimated using a simple $KS$-test 
which determines the level at which the 
distribution of LBG fluxes is inconsistent with being drawn randomly from the distribution of pixel 
values 
in the sub-mm map.  The $KS$ 
probabilities are given in Table \ref{tbl-2}.  There are detections at approximately 2$\sigma$ (KS 
probability 
$\simeq$90\%) for the two CFDF samples while the  Steidel 
LBG sample is consistent with zero mean flux.  The distributions of sub-mm fluxes values for the three 
LBG samples are shown in Figure 1.  \

We have also investigated the effect on our analysis if we do not remove any sources from the maps.  
In 
this case, we obtain the following results 
for the three samples: -0.171 $\pm$ 0.263 mJy for CFDF-03, 0.760 $\pm$ 0.206 mJy for CFDF-14, 
and -0.021 $\pm$ 0.110 mJy for 
Steidel-14.  For the  14-hour CFDF sample,  we now have a  highly significant result while for the 
other 
two fields the mean flux has decreased.  
It may seem counter-intuitive that the coadded flux could 
actually decrease if the bright sub-mm sources are not removed but this is simply a result of the beam 
profile.  The chopping technique creates two 
{\it negative} sources for every positive source (offset by 30 arcsecs in each direction of RA) and so if 
these are not removed they can lead to a 
decrease in the coadded flux (especially if the LBGs are clustered around the bright sub-mm sources). 
Of 
course, not removing bright sources can 
also lead to an increase in the coadded flux, as in the case for the 14-hour CFDF sample, if a 
significant 
number of the LBGs are close to the peak 
of 
the positive beam (although still with a $>$ 8 arcsecs offset).    \

\begin{deluxetable}{cccc}
\tabletypesize{\scriptsize}
\tablecaption{Lyman-Break Galaxies with S$_{850{\mu}m}>$ 3 {\rm mJy} \label{tbl-1}}
\tablewidth{0pt}
\tablehead{
\colhead{Source Name} & \colhead{S$_{850{\mu}m}$ (mJy)} & \colhead{LBG Name} & \colhead{P 
probabilities}
}

\startdata
CFRS14-6 & 4.2  & CFDF 64601 & 0.019 \\
CFRS14-7 & 3.2 & Steidel West2-MD13 & 0.096 \\
CFRS14-8 & 3.4 & Steidel MMD75 & 0.081 \\
CFRS14-10 & 3.5 & Steidel MMD63 & 0.072 \\

\enddata
\end{deluxetable}


\begin{deluxetable}{cccc}
\tabletypesize{\scriptsize}
\tablecaption{Submillimeter Flux of Lyman-Break Galaxies\label{tbl-2}}
\tablewidth{0pt}
\tablehead{
\colhead{LBG Sample} & \colhead{Mean Flux}   & \colhead{KS significance} &\colhead{Mean Flux} \\
\colhead{} & \colhead{(sub-mm sources removed)}  & \colhead{} & \colhead{(no sources removed)}
}
\startdata
CFRS03+00 CFDF & 0.414 $\pm$ 0.263   & 94.4 \% & -0.171 $\pm$ 0.263 \\
CFRS14+52 CFDF & 0.382 $\pm$ 0.206   & 88.7 \%  & 0.760 $\pm$ 0.206 \\
CFRS14+52 Steidel & 0.0141 $\pm$ 0.110 &  18.4 \% & -0.021 $\pm$ 0.110 \\

\enddata
\end{deluxetable}


\begin{figure}
\figurenum{1(a)}
\epsscale{1.0}
\plotone{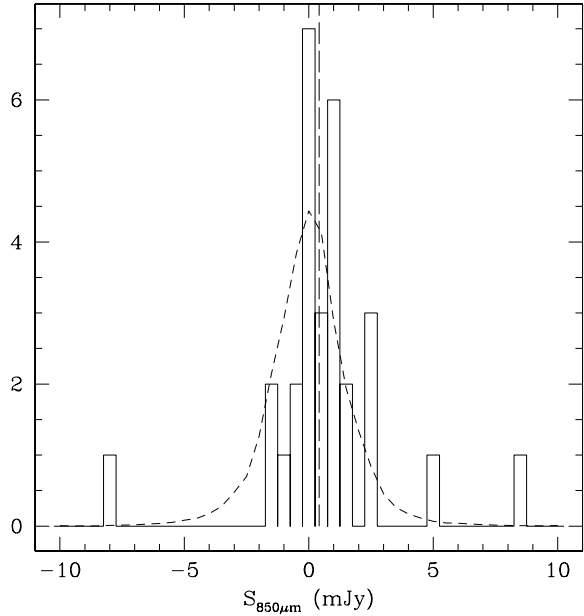}
\caption{Sub-mm flux distribution for the Lyman-break galaxies in the 3-hour sample. Overlaid is the 
distribution of all the pixel values in the 
3-hour submillimeter map (with all sources removed) to illustrate the flux levels at which the LBG flux 
distribution deviates from that of the 
sub-mm map.  The weighted mean of the sample is also shown (vertical dashed line).  The sub-mm 
3-hour 
field had poorer weather than the 
14-hour field and is therefore noisier.  The 2 measurements at $>$5 mJy and 1 at $<$ -5 mJy are in 
extremely noisy regions of the image (with $S/N < $ 3$\sigma$) and  as 
such have a very low weighting factor in the mean.  }
\end{figure}


\begin{figure}
\figurenum{1(b)}
\epsscale{1.0}
\plotone{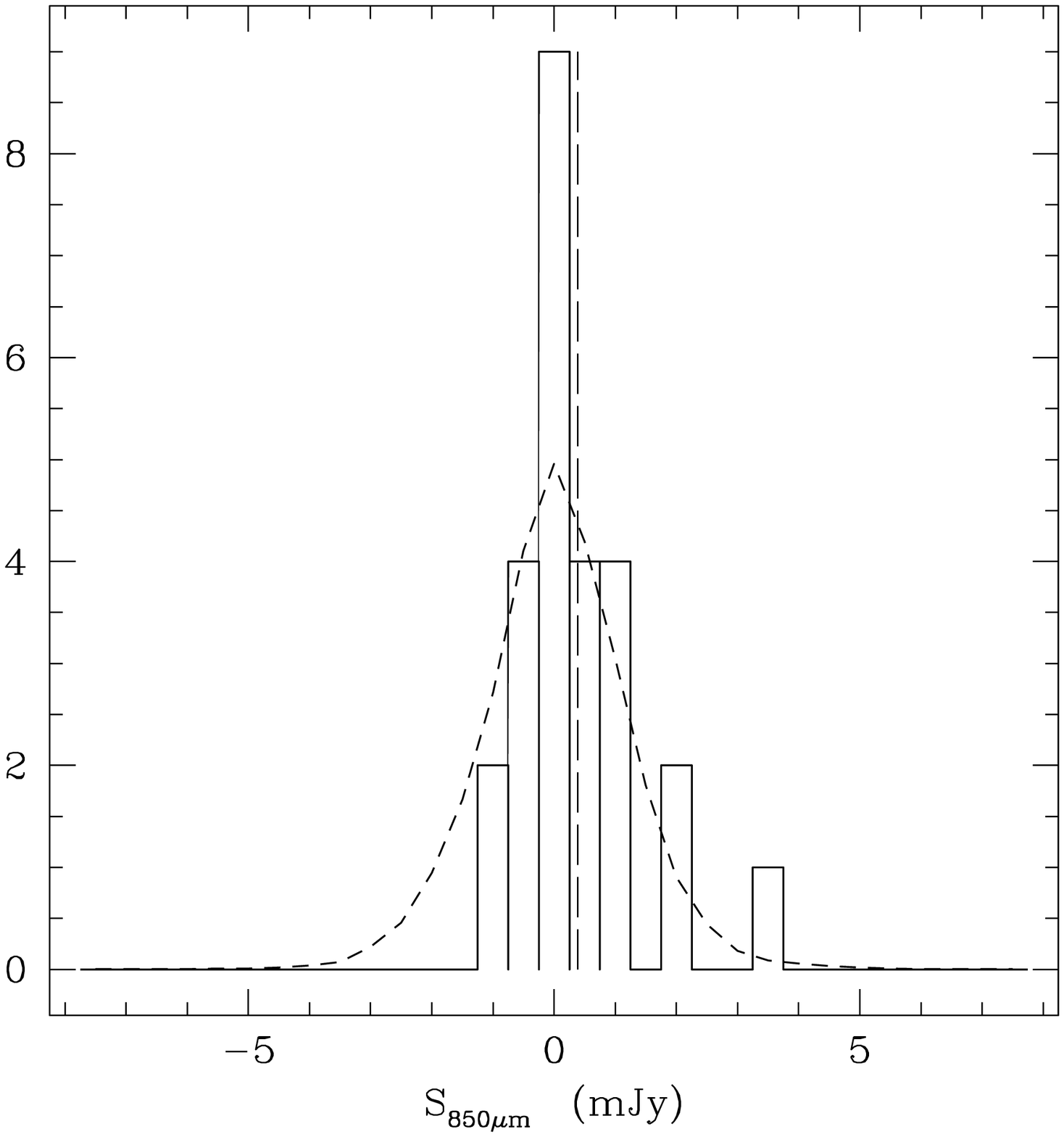}
\caption{Same as Fig. 1(a) but for  the Lyman-break galaxies in the 14-hour CFDF sample.  Overlaid is 
the distribution of all the pixel values in 
the 14-hour submillimeter map (with all sources but the one source in Table 2 removed).  The weighted 
mean (vertical line) is also shown.}
\end{figure}


\begin{figure}
\figurenum{1(c)}
\epsscale{1.0}
\plotone{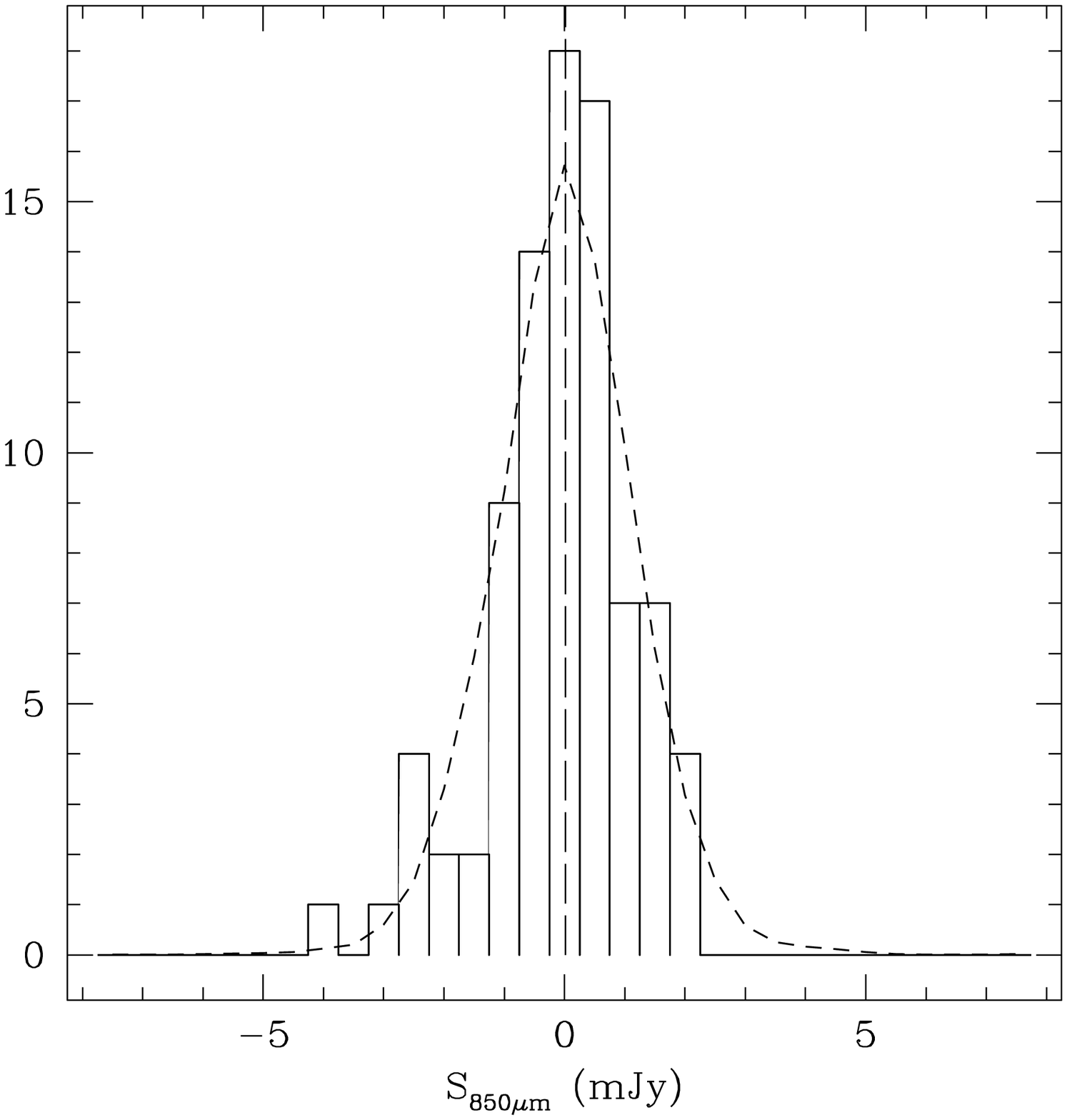}
\caption{Same as Fig. 1(a) but for the Lyman-break galaxies in the 14-hour Steidel sample. Overlaid is 
the 
distribution of all the pixel values in the 
14-hour submillimeter map (with all sources but the three listed in Table 2 removed).  For consistency 
the 
flux for each object was taken to be the 
flux at the position of the LBG, including the three LBGs possibly associated with SCUBA objects at 
$>$ 
3$\sigma$.  These three objects have 
sub-mm fluxes (measured at their sub-mm location) of $<$ 3.5 mJy and therefore because of the 
offset 
between the LBG and SCUBA positions 
have recovered fluxes of $\sim$2 mJy.   }
\end{figure}


\subsection{The Lyman-Break Galaxy Contribution to the Submillimeter Background}

The contribution from LBGs to the FIR/sub-mm background is still an open question and one which 
can 
be 
addressed by these data.  The bright 
sub-mm selected  sources ($S_{850{\mu}m} >$ 3 mJy)  produce 20-30\% of the background energy at 
850$\mu$m \citep{bar98,bar99,hug98,bla99,eal00,cow02,sma02}, and therefore if 
the LBGs are  the same population  as the $S_{850{\mu}m} <$ 3 mJy objects they  could be responsible 
for 
the remaining 70-80\%.  \citet{adel00} 
estimate  that $UV$-selected galaxies with  $m_{1600}<$ 27 could easily have produced $\sim$75\% 
of 
the sub-mm background at 850$\mu$m. 
However, in this picture the bulk of this energy is emitted by LBGs too  faint to be included in our 
optical 
sample and at sub-mm fluxes too faint to 
be statistically detected using our method. \citet{cha00}, who targeted brighter LBGs  ($\mathcal{R} < 
$24.5) which are expected to be stronger 
sub-mm emitters, concluded that these objects produce a negligible contribution to the background. \

  Given that none of our three LBG samples has been strongly detected as a population at 850$\mu$m  
we 
estimate upper-limits 
to the background contribution that these galaxies make at this wavelength.   Given the 1$\sigma$ 
standard error of the coadded flux 
measurements in  Table 1 we can estimate a 3 $\sigma$ upper limit for the background contribution.  
For 
the CFDF samples, in the redshift range 
2.9 $ < z < $ 3.5  the 3-hour field contributes $< 3.2\%$ to the background and the 14-hour field 
contributes $< 2.8\%$.  For the Steidel et al 
sample, which is selected over a larger redshift range of 2.4 $< z < $ 3.4, the contribution is $< 
5.1\%$. 
   \

However, because of the negative 
$K$-correction, the observed 850$\mu$m flux is constant with redshift (for approximately 0.5 $ < z < 
$ 
6, depending on cosmology)  and the 
contribution of LBGs to the submillimeter  background 
is coming from a wide range of redshifts, not just the tight range over which the Lyman-break 
technique 
selects galaxies.  To determine the total 
contribution we must include the sub-mm flux from LBGs outside the selected redshift range.  To do 
this 
we assume that LBGs have a constant 
comoving number density between redshifts 1 and 5, and then, by integrating the comoving volume 
increment ($dV_c/dz$) over this range, we 
can scale the background contribution accordingly.  Thus, upper limits on the total contribution to the 
background at 850$\mu$m becomes $<$ 
20\% for CFDF-03, $<$ 16\% for CFDF-14 and $<$ 19\% for the deeper Steidel et al sample in the 
14-hour field.

\subsection{ Lyman Break Galaxies Identified with SCUBA Sources}

The optical and near-IR counterparts to the $>$ 3 mJy SCUBA sources are discussed in detail in Webb 
et 
al. (2001b, in preparation) and Clements 
et al. (2001, in preparation), along with the 
identification 
process.  There are four SCUBA sources whose best identification is a Lyman-break galaxy, three in 
Steidel et al's  sample and  one from  
CFDF-14.  \

It is  interesting to note that the three  SCUBA sources identified with Steidel et al objects are part of 
a 
chain-like structure in the 
sub-mm map and are all within 40  arcseconds of  each other.  The remaining source, associated with 
an 
object in the CFDF-14 sample (but not 
present in the Steidel et al sample) is  also found in  the same general area approximately 1 arcmin to 
the 
west.  \

West2-MD13, which is identified 
with  14.7, is the only object of the three possible Steidel LBG identifications with a spectroscopic 
redshift.  This object is a quasar 
which lies in the largest single over-density within the redshift range to which the Steidel et al survey 
was 
sensitive. There is reason to expect that 
SCUBA  sources might be associated with over-dense regions in the early universe, (see Discussion). 
For 
example,  \citet{cha01b} observed an over-density of LBGs at $z$=3.09 \citep{ste98} with SCUBA 
and 
detected a correspondingly high surface density  of sub-mm  sources.

\section{Dust Properties of Lyman-Break Galaxies}

We have obtained an  upper limit (2$\sigma$)  of $\simeq$0.4 mJy on the average 850$\mu$m flux
of LBGs for both CFDF samples. We converted this to an upper
limit on the dust mass using the following formula \citep{hil83}:
\smallskip
$$
M_d = { (1 + z) D^2 S_{obs} \over \kappa_d(\nu_{em}) B(\nu_{em},T_d) } \eqno(1)
$$
\smallskip
\noindent where $(1+z)D$ is the luminosity-distance and $S_{obs}$ is the flux.
We assumed that all the LBGs are at a redshift
of 3, which means that the emitted frequency, $\nu_{em}$, is $\rm 1.41
\times 10^{12}$ Hz, equivalent to 212.5$\mu$m.
We assumed that the dust mass opacity
coefficient, $\kappa_d(\nu)$, has a value of 0.077 m$^2$ kg$^{-1}$
at 850$\mu$m and extrapolated this to the shorter wavelength 
using the formula $\kappa_d(\nu) \propto \nu^{\beta}$ with
a dust-emissivity index, $\beta$, of 2, for which there
is now strong evidence (\citet{dun01} and references therein). To compare this upper limit with the dust 
masses of nearby
galaxies, we also calculated dust masses for the 104 galaxies in the
SCUBA Local Universe and Galaxy Survey (SLUGS, \citet{dun00}) using the
same formula. This survey contains a variety of galaxy types, from ULIRGs like Arp 220
to galaxies which are much more typical of the normal galaxy population.
To avoid the uncertainty in $\beta$ making the comparison
between the high and low-redshift galaxies dubious, we used the multi-wavelength
data that exists for the SLUGS galaxies to estimate the fluxes of the 
galaxies at the same frequency used for the LBGs. Therefore, even if the value of $\beta$ we have 
assumed
to calculate the dust mass opacity coefficient is incorrect, the
relative dust masses of the LB and SLUGS galaxies will be correct.
The value of the dust mass opacity coefficient at any frequency
is still quite uncertain \citep{alt00}, and so 
our absolute dust masses may not be correct,
but again our comparison of the dust masses of low- and high-redshift
galaxies will be valid, as long as the properties of dust are not
radically different at low- and high redshift \citep{eal96}.

Since the emitted frequency for the LBGs is fairly close to
the peak of the dust spectral energy distribution, our dust mass estimates
are much more sensitive to the assumed dust temperature than at lower
frequencies. \citet{dun01} have recently shown that, for the
SLUGS galaxies, the mass-weighted
temperature of the dust (the correct one to use in the formula above)
is relatively low, even for extreme ULIRG's like Arp 220. They derive
an average mass-weighted temperature for the sample of $\rm 21.3\pm0.5$ K.
We have used this value to derive the dust masses for the LBGs and
the individual values of mass-weighted temperature for the SLUGS galaxies. Fig. 2 shows the results 
for 
the three different cosmologies.  
Although the mass limit for the LBGs depends quite strongly on the cosmological model, it is clear that 
their dust masses must be no larger 
than those of nearby galaxies.  The significance of these dust masses will be discussed by Dunne et al. 
(in preparation). 

\begin{figure}
\figurenum{2}
\epsscale{.7}
\plotone{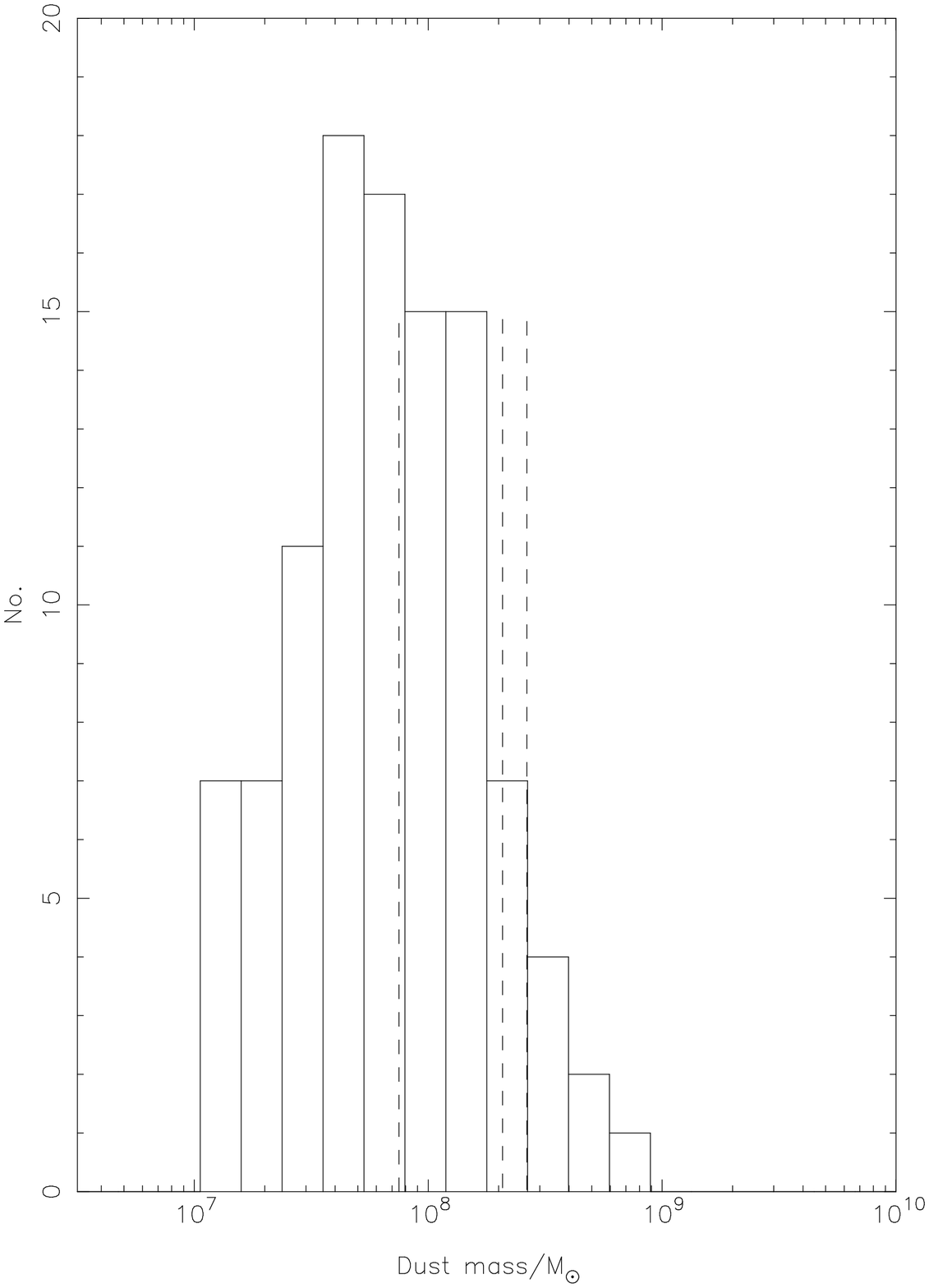}
\caption{Histogram of the dust masses of galaxies from the SCUBA Local
Universe and Galaxy Survey \citep{dun00}, superimposed with vertical
dashed lines showing the upper limit on dust mass for the LB galaxies obtained
in this paper. The three lines correspond to different cosmologies: from
left to right, ($\Omega_0=1$, $\Lambda=0$), ($\Omega_0=0.2$, $\Lambda=0.8$), ($\Omega_0=0$, 
$\Lambda=0$).}

\end{figure}


\section{The Cross-Correlation Function Between SCUBA Sources and Lyman-Break Galaxies}

We  measured the SCUBA-LBG angular cross-correlation function in each field with a separate 
analysis 
for each sample in  the 14-hour field.
Figure 3 shows the positions of the LBGs and SCUBA sources for all three LBG samples. First, we 
calculated  the angular separations of all the 
$N_s \times N_l$ SCUBA-LBG pairs in each field  and then divided these
into bins of angular distance. This procedure gives the basic function from the data, $N_{sl}(\theta_i)$ 
or 
$LS$.

We then generated a list of random galaxy positions for both the SCUBA sources and LBG galaxies
taking account of sensitivity variations in the images.
Since we have two different sets of sources - SCUBA sources and
LBGs - it is necessary to take account of the selection
effects for both sets of sources. The Lyman-break catalogues are
subsets of larger catalogues drawn from a much larger area of sky than
the SCUBA fields, and we generated lists of 5000 random positions for
Lyman-break galaxies on the assumption that the sensitivity of all
the catalogues is constant over the SCUBA fields. \

The sensitivity
of each SCUBA field is, however, very inhomogeneous \citep{eal00,web02}.
To generate
lists of random positions for the SCUBA sources that take account
of the variation in sensitivity of each SCUBA image, we adopted the
following procedure. Using the
best-fit submillimeter source counts
from our survey \citep{eal00,web02}  we produced lists of artificial random SCUBA sources.
For each source, we  produced a random 
position  {\it on the assumption that there is no
variation in sensitivity across the fields.} We then used the
noise images for each field  \citep{eal00,web02}  to determine whether
each artificial source would have been detected in our survey. In this
way we produced a list of 5000 artificial SCUBA sources.

We calculated the angular separations of the real SCUBA
sources and the $N_{r_l}$ artificial LBGs, giving
the function $SR_L$. 
We did likewise with the real Lyman-break galaxies and artificial SCUBA sources ($LR_S$),
and finally the artificial SCUBA sources and artificial Lyman-break galaxies,
giving the function
$R_L R_S$.  These functions must all be normalized to the same number of pairs as used to calculate 
$LS$.

We used   two possible estimates of $w(\theta)$. The first of these is the Landy \& Szalay formalism 
\citep{lan93}:

\smallskip

$$
w(\theta) = {LS - (LR_S + SR_L) + R_L R_S \over R_L R_S} \eqno(2)
$$
\smallskip

\noindent and the second is the  Hamilton formula \citep{ham93}

\smallskip

$$
 w(\theta) = {LS  \times R_LR_S \over LR_S \times SR_L} -1\eqno(3)
$$

\noindent Both functions gave comparable results.  \

The final complication is
that of the ``integral constraint''. If $w(\theta)$ is estimated
from an image, the integral 
\smallskip
$$
{ 1 \over \Omega^2} \int \int w_{est}(\theta) d\Omega_1 d\Omega_2 \eqno(4)
$$
\smallskip
\noindent will necessarily be  zero, even though the
same integral of the true correlation function will not be zero
for any realistic image size \citep{gro77}. 
As in \citet{roc99}, we assumed that the observed
angular correlation function is given by:
\smallskip
$$
w(\theta) = A (\theta^{-0.8} - C) \eqno(5)
$$
\smallskip
\noindent $C$ can then be calculated from
\smallskip
$$
C = { \sum \theta^{-0.8}_{ij} \over N_{r_l}N_{r_s} } \eqno(6)
$$
\smallskip
\noindent in which $\theta_{ij}$ is the angular distance between
the i'th artificial Lyman-break galaxy and the j'th artificial SCUBA
source.  For the 14-hour field this was calculated to be 0.0174 and for the 3-hour field, 0.0158.

\begin{figure}
\figurenum{3(a)}
\epsscale{1.0}
\plotone{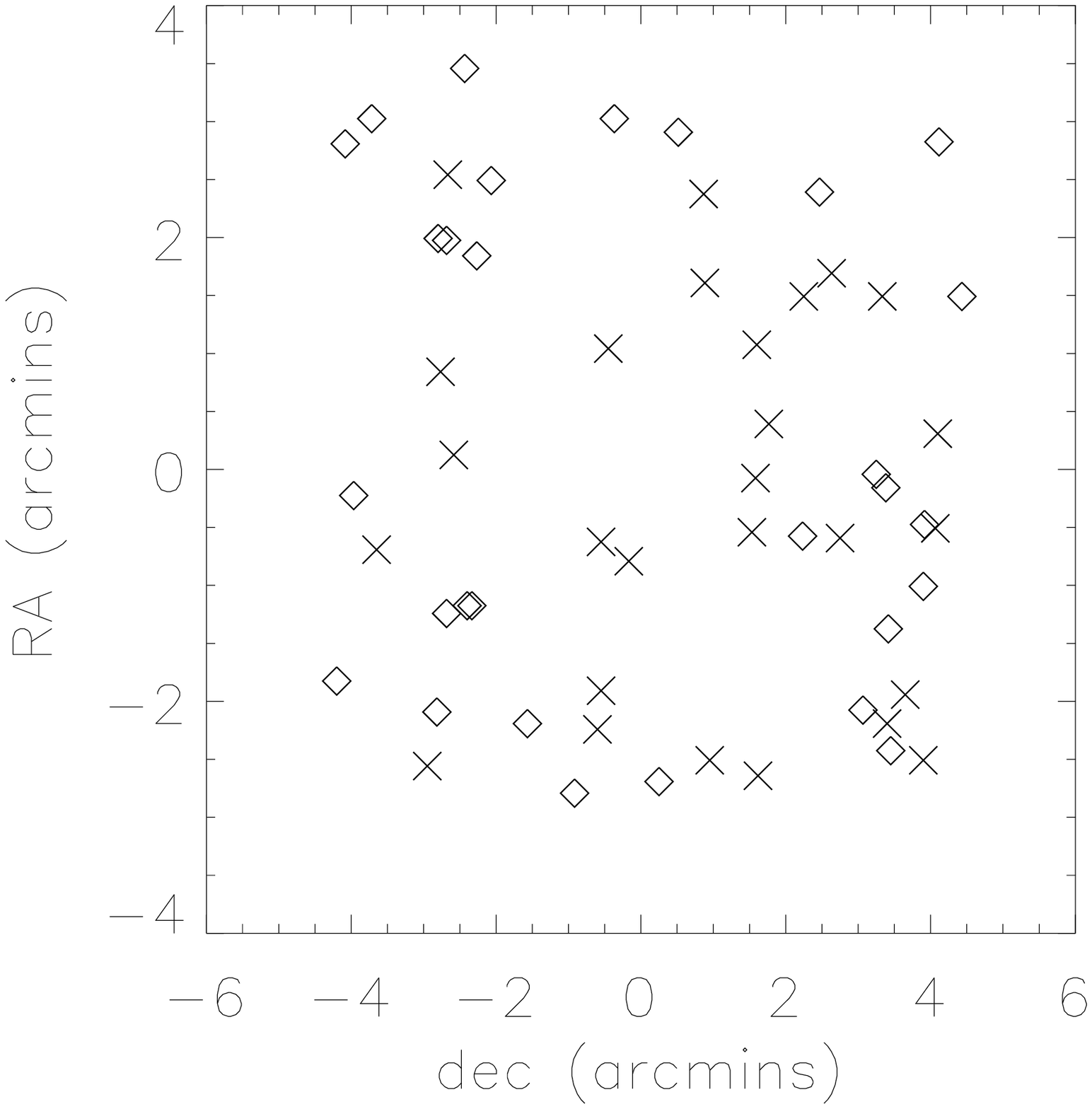}
\caption{This figure shows the relative positions of the SCUBA sources and LBGs in the CUDSS-3 
field.  
Diamonds correspond to  LBG's and  
crosses to  SCUBA sources. The sensitivity of the SCUBA map varies substantially over the entire field 
while the LBGs are considered to be 
uniformly selected.  }
\end{figure}


\begin{figure}
\figurenum{3(b)}
\epsscale{1.0}
\plotone{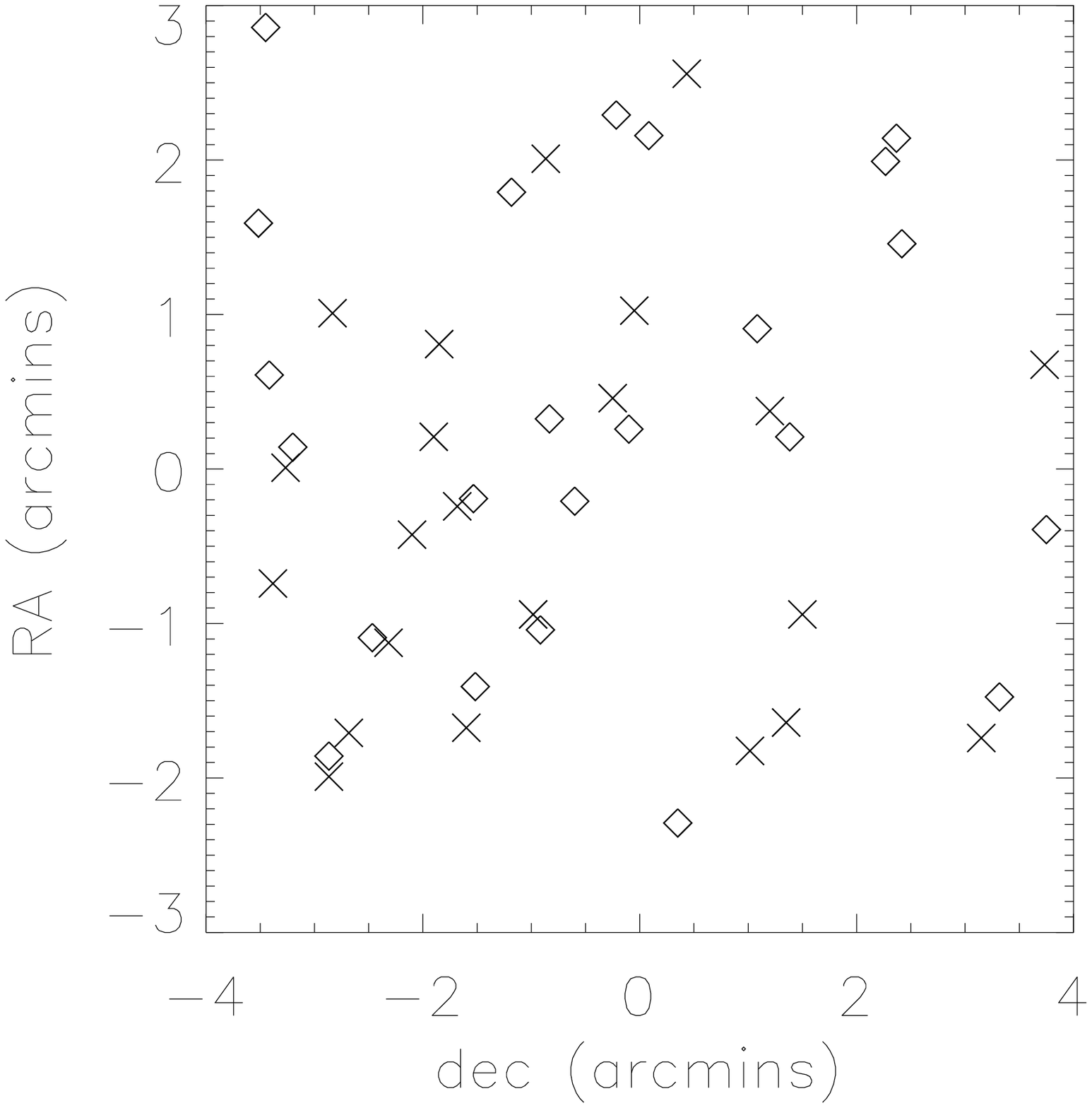}
\caption{Same as figure 3(a) but for the CUDSS-14 field and the CFDF 14-hour field LBG sample.}
\end{figure}


\begin{figure}
\figurenum{3(c)}
\epsscale{1.1}
\plotone{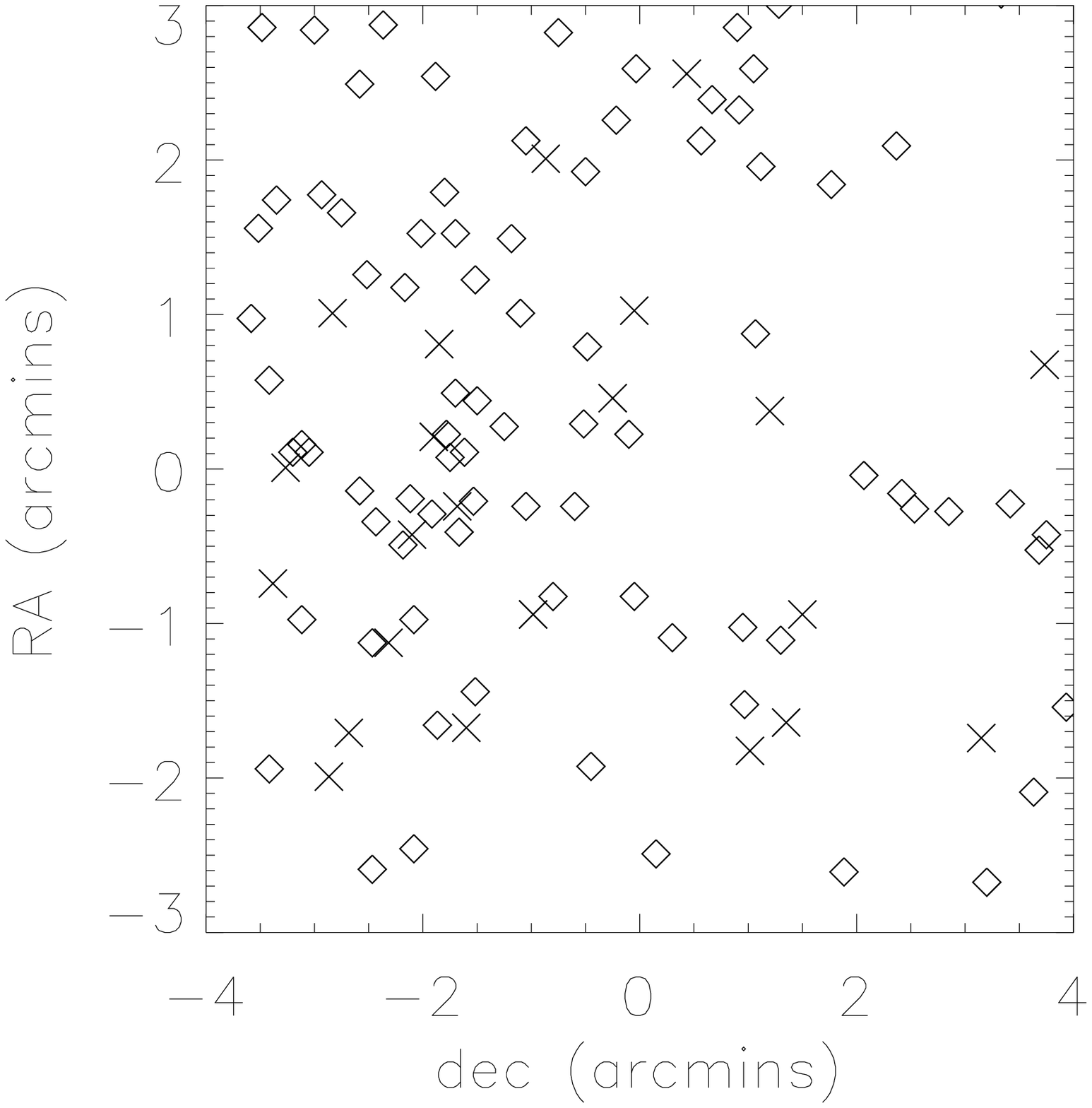}
\caption{Same as for figure 3(a) and (b) but for the CUDSS-14 field and the Steidel et al 14-hour 
field 
LBG sample.}
\end{figure}

Fig. 4 shows our estimates of w($\theta$) for both fields.   To estimate our errors we used the 
bootstrap 
method as outlined by  \citet{bar84}.
We then fitted  Equation 5 using these values of C with the ${\omega}(\theta)$ results from each field 
and  
determined a cross-clustering amplitude of 5.2 
$\pm$ 2.9
arcsec$^{0.8}$ for the Steidel 14-hour sample, 1.1  $\pm$ 4.4 arcsec$^{0.8}$ for CFDF-14 and and 
2.3 
$\pm$ 3.8 arcsec$^{0.8}$ for  CFDF-3, 
where the errors are simply errors in the chi-squared fit.  At separations of 10 arcsecs these 
amplitudes 
correspond to ${\omega}(\theta)$ = 0.82  
$\pm$ 0.46 for the Steidel sample, ${\omega}(\theta)$ = 0.17 $\pm$ 0.70 for CFDF-14 and 
${\omega}(\theta)$ = 0.36 $\pm$ 0.60 for CFDF-3. 
Clearly, the result from the Steidel et al sample in the 14-hour field, which is the largest 
sample of the three,  is the most (only) significant result, and is quite large. 

In order to investigate the possibility that the 14-hour field is an 
unusually clustered region we calculated the auto-correlation function of the Steidel et al LBGs.  The 
result 
is consistent 
with the results found in \citet{gia98} using many fields over a larger area. For the brighter CFDF 
sample 
Foucaud et al. (in preparation) 
calculated an amplitude of 4.0
$\pm$ 0.7 
arcsec$^{0.8}$ for CFDF-14 which is slightly lower than the average of the three CFDF fields 
(CFDF-14, CFDF-3 and CFDF-22) of 4.7 $\pm$ 
1.2 arcsec$^{0.8}$. Therefore, the 14-hour field does not appear to be exceptional.   The amplitude of 
the 
SCUBA-LBG cross-correlation function for 
CFDF-3 is comparable to the amplitude of the auto-correlation function of the brighter  LBGs in the 
CFDF sample.   \

This analysis was performed assuming that the LBGs and SCUBA sources were not the same object 
and 
so those objects within 8 arcsecs of each 
other were included in the correlation signal.  To be certain that these few objects were not biasing the 
clustering signal we redid the analysis, with 
the added restriction of ${\theta}>$ 8 arcsecs.  We find the following amplitudes: for the Steidel 
sample, 
6.7 $\pm$ 1.9 arcsec$^{0.8}$, for 
CFDF-14, 0.4 $\pm$ 3.4 arcsec$^{0.8}$ and for CFDF-03, 3.8 $\pm$ 3.9 arcsec$^{0.8}$.  It does 
not 
appear, therefore, that the possible 
LBG-SCUBA associations are strongly affecting the correlation results.  \
 
In this analysis we have used the complete CUDSS SCUBA catalogue of 50 sources.  As this sample 
includes all sources detected above 3$\sigma$ one might worry that a large number of spurious sources 
are contaminating the clustering analysis.  However, in \citet{eal00} and \citet{web02} we discuss the 
number of expected spurious sources in the sample, using Gaussian statistical arguments and 
arguments based on an analysis of the noise in the SCUBA maps.  We have concluded that about 2-3 
sources in each field are spurious, or a total of 4-6 sources in the combined sample (or approximately 
10\%).  At this level these sources are not expected to significantly alter the cross-clustering analysis.  

However, to check this, we have repeated the clustering measurement using a subset of the 26 
SCUBA sources in 
our catalogue which were detected at $\geq$ 3.5$\sigma$.  For a larger sample size one might expect 
the clustering amplitude to increase with the removal of the spurious sources in the sample, however, in 
doing so we are substantially decreasing our sample size (as we must remove all $<$ 3.5$\sigma$ 
sources, not just the 10\% which are spurious).  We measure the following clustering amplitudes: 
$A$=2.7$\pm$5.2 arcsec$^{0.8}$ for the 3-hour field,     $A$=1.2$\pm$5.4 arcsec$^{0.8}$ for the 
14-hour field and CFDF-14, and $A$=5.1$\pm$5.5 arcsec$^{0.8}$ for the 14-hour field and the 
Steidel et al. sample, where the errors have again been estimated using the bootstrap method.  The 
measured values are essentially the same as the amplitudes measured for the entire $\geq$3.0$\sigma$ 
sample, though the uncertainties have increased substantially because of the decrease in sample size.  
Hence, it does not appear that the analysis has been contaminated by spurious sources.   

\begin{figure}
\figurenum{4}
\epsscale{1.0}
\plotone{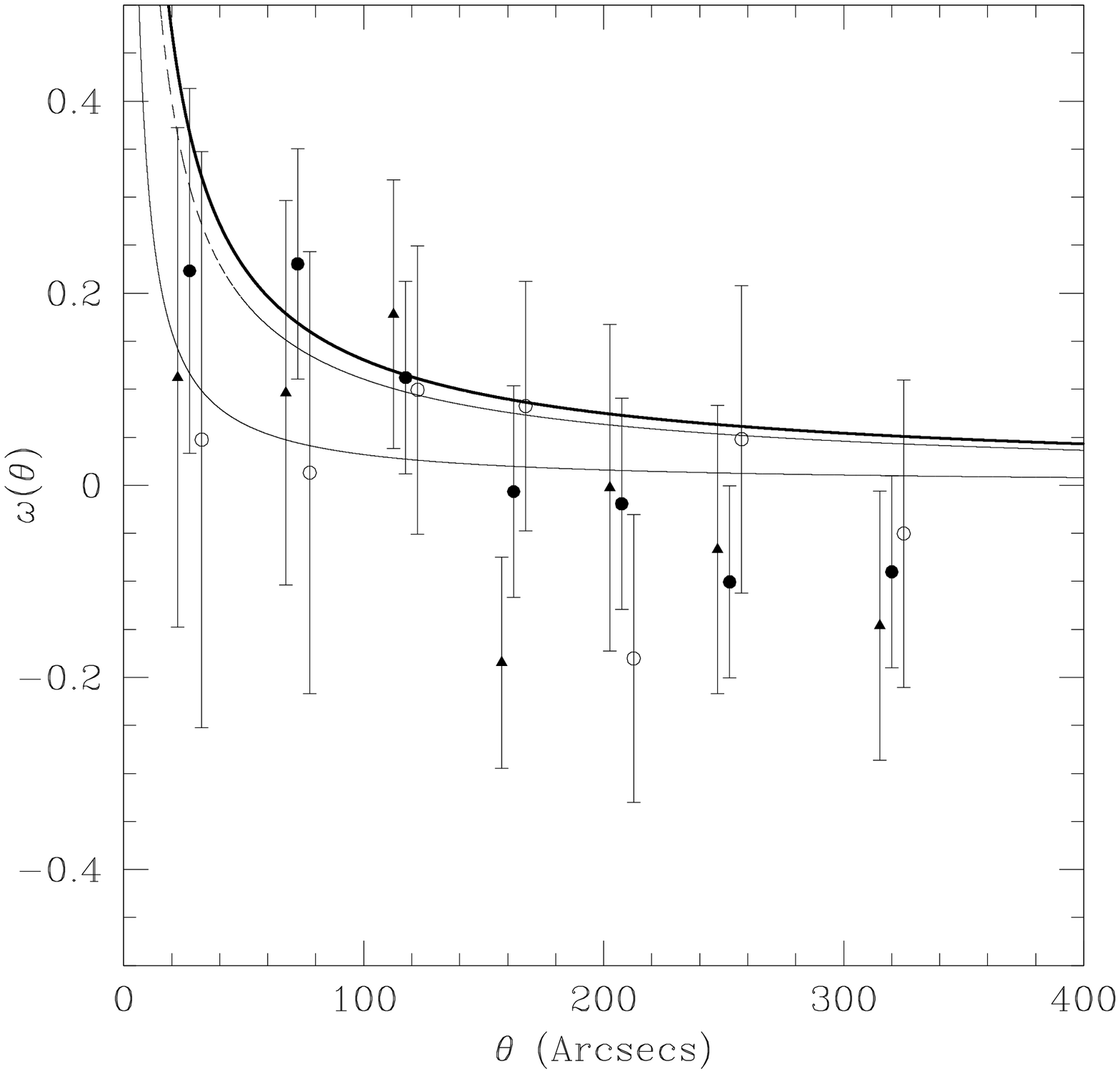}
\caption{Angular cross-correlation function (SCUBA-LBG) for the three Lyman-break samples.  The 
triangles are for the 3-hour field, the filled circles are for the 
Steidel  et al. sample of the 14-hour field and the open circles are for the CFDF sample of the 14-hour 
field. The upper, bold solid line   is the 
LBG-SCUBA cross-correlation function measured in this paper (not including the integral constraint)  
for 
the 
Steidel et al sample.  This fit used only the inner four points since beyond about 150 arcseconds edge 
effects become important. The middle dotted-line corresponds to the auto-correlation function of the 
bright 
(S$_{850{\mu}m} >$ 3 mJy) SCUBA sources from \citet{web02}. The lower solid-line is the 
auto-correlation 
function
for  the Lyman-break galaxies from \citet{gia01}. }
\end{figure}

\section{Discussion}

\subsection{Submillimeter Flux of Lyman-Break Galaxies}

We  detect, at a level of $\sim$2$\sigma$, a mean sub-mm flux from the two CFDF samples of LBGs 
but 
we do not detect any flux from the 
Steidel sample.  Recalling that the CFDF 
14-hour  LBGs are an optically bright subset of the Steidel LBGs this could be taken as evidence of a 
positive correlation between observed 
optical and  sub-mm flux.  However, a plot of $I$-band magnitude versus recovered sub-mm flux  
shows 
no such correlation (Figure 5).  Similarly, 
no correlation is  present when $\mathcal{R}$-band magnitude for the  Steidel sample is plotted versus 
sub-mm flux.    In addition, there is no 
statistical 
detection of the  sub-mm flux of a brighter sub-sample ($\mathcal{R} < $ 25 ) of the Steidel et al list 
(which 
agrees with the CFDF-14 catalogue at 
approximately the  70\% level).   \

On closer inspection we find that the 
CFDF-14 measurement is completely dominated by 
the detection of  one LBG  and when this object is removed from the list a positive mean flux is 
detected 
only at the 1$\sigma$ level.  Although 
there are three  detections of LBG's at $>$3$\sigma$ in the Steidel list  they do not lift the mean flux 
of 
the entire 86 galaxy sample to a 
significant level.  In CUDSS-3, where no LBG was detected above 3$\sigma$  the positive detection 
of 
flux is not  due to 
any single object. \

It is not entirely surprising that we see no correlation between the sub-mm  and observed optical flux.  
Many authors 
\citep{sha01,adel00} have claimed  that galaxies with higher intrinsic $UV$ luminosities (after correcting 
for reddening) contain larger dust masses.  
However,  this does not necessarily translate into a correlation 
between {\it observed} optical and sub-mm flux.  Indeed, as with our results, \citet{sha01} found no 
correlation between dust content and 
observed optical flux.  \

Two other groups have attempted to directly measure the sub-mm flux of high-redshift star forming 
galaxies. \citet{cha00} carried out a targeted 
study of  8 Lyman-break galaxies, selected to have high  $UV$-derived star formation rates and they 
obtain a  result similar to our own. Although 
one object was detected at  $>$3$\sigma$, no flux was detected statistically from the sample as a 
whole 
once this object was 
excluded.  However, in a recent conference  proceeding  \citep{cha01a}, with a larger sample of 33 red 
and 
high-star forming LBGs, they report 
a  statistical detection of  S$_{850}$=0.6 $\pm$0.2 mJy  with marginal detections for several individual 
red 
LBGs. \

\citet{pea00} followed a similar approach to this paper for  starburst galaxies in the  Hubble  Deep Field 
and 
detected a mean flux of 0.2 $\pm$ 0.04 
mJy (for a SFR of 1 $h^{-2}$M$_{\odot}$yr$^{-1}$) with the mean flux increasing with SFR.   To 
directly 
compare this result with our own we 
must first convert our statistical measurement to units of submillimeter flux per unit star formation rate. 
To 
determine the UV-estimated star 
formation rates for the LBGs in our three samples we follow the method outlined in \citet{pea00} and 
\citet{mad96}.  Approximately 60\% of the 
Steidel et al.~LBGs in our field have spectroscopic redshifts.  For those without spectroscopic 
redshifts, 
and for the CFDF LBGs, we assumed a 
redshift of $z$=3. We chose a flat,  $\Omega_{\Lambda}$=0.7 cosmology. We find the following:

\

\noindent for the Steidel et al. 14-hr sample
$$
S_{850{\mu}m}\mbox{/mJy} = 0.015\pm 0.022 {\mbox{SFR}\over 
h^{-2}\mbox{M}_{\odot}\mbox{yr}^{-1} 
 } 
\eqno(9) \
$$
\smallskip
for the CFDF 14-hr sample
$$
S_{850{\mu}m}\mbox{/mJy} = 0.065\pm 0.034 {\mbox{SFR}\over 
h^{-2}\mbox{M}_{\odot}\mbox{yr}^{-1} 
 } 
\eqno(10) \
$$
\smallskip
for the CFDF 3-hr sample
$$
S_{850{\mu}m}\mbox{/mJy} = 0.13\pm 0.06 {\mbox{SFR}\over h^{-2}\mbox{M}_{\odot}\mbox{yr}^{-1}  } 
 \eqno(11) \
$$
\smallskip

Of our three LBG samples only the statistical measurement from CFDF-03 is consistent with the 
results 
of \citet{pea00}.  The Steidel et al. sample 
and the CFDF-14 sample both have much lower detections.  \citet{pea00}  converted the results of 
\citet{cha00} to these units and found:
\smallskip
$$
S_{850{\mu}m}\mbox{/mJy} = 0.13\pm 0.14 {\mbox{SFR}\over h^{-2}\mbox{M}_{\odot}\mbox{yr}^{-1}  } 
 \eqno(12) \
$$
\smallskip
\noindent which is consistent with the \citet{pea00} result, although also consistent with no detection 
of 
submillimeter flux. It appears that most 
LBGs are not strong sub-mm emitters, that is with 0.3 $< S_{850{\mu}m} < $ 3 mJy, and therefore the 
strength of a statistical submillimeter 
detection may depend strongly on the properties of the specific LBG sample observed.   \

In figure 5 we plot the mean 850$\mu$m flux as a function of $UV$-estimated star formation rate for 
all 
three LBG samples.   Though, to the eye,  somewhat suggestive of a rise, these data are consistent 
with 
no 
correlation  between the star formation rate and the submillimeter flux.  \ 

As discussed in \S 4 there are four Lyman-break galaxies which are possibly identified with a SCUBA 
source at $>$3$\sigma$, one from the 
CFDF 14-hour sample and three from the Steidel et al sample.   There are very few  Lyman-break or 
high-redshift star forming galaxies 
(from all studies)  which are known to be sub-mm bright and therefore conclusions regarding any 
unifying 
properties are difficult.  There is some 
suggestion however that these objects have extremely red colors compared to the average for the 
population \citep{cha00,cha01a}. As 
discussed 
in \citet{sha01}  the more intrinsically luminous LBGs appear to be dustier, with redder optical colours, 
and 
should therefore be brighter in the 
sub-mm.   However, we see no such trend with the four LBGs identified with bright SCUBA objects in 
our sample 
although we have very limited photometric information, particularly in the near-infrared,  on these faint 
objects. \

\begin{figure}
\figurenum{5}
\epsscale{1.0}
\plotone{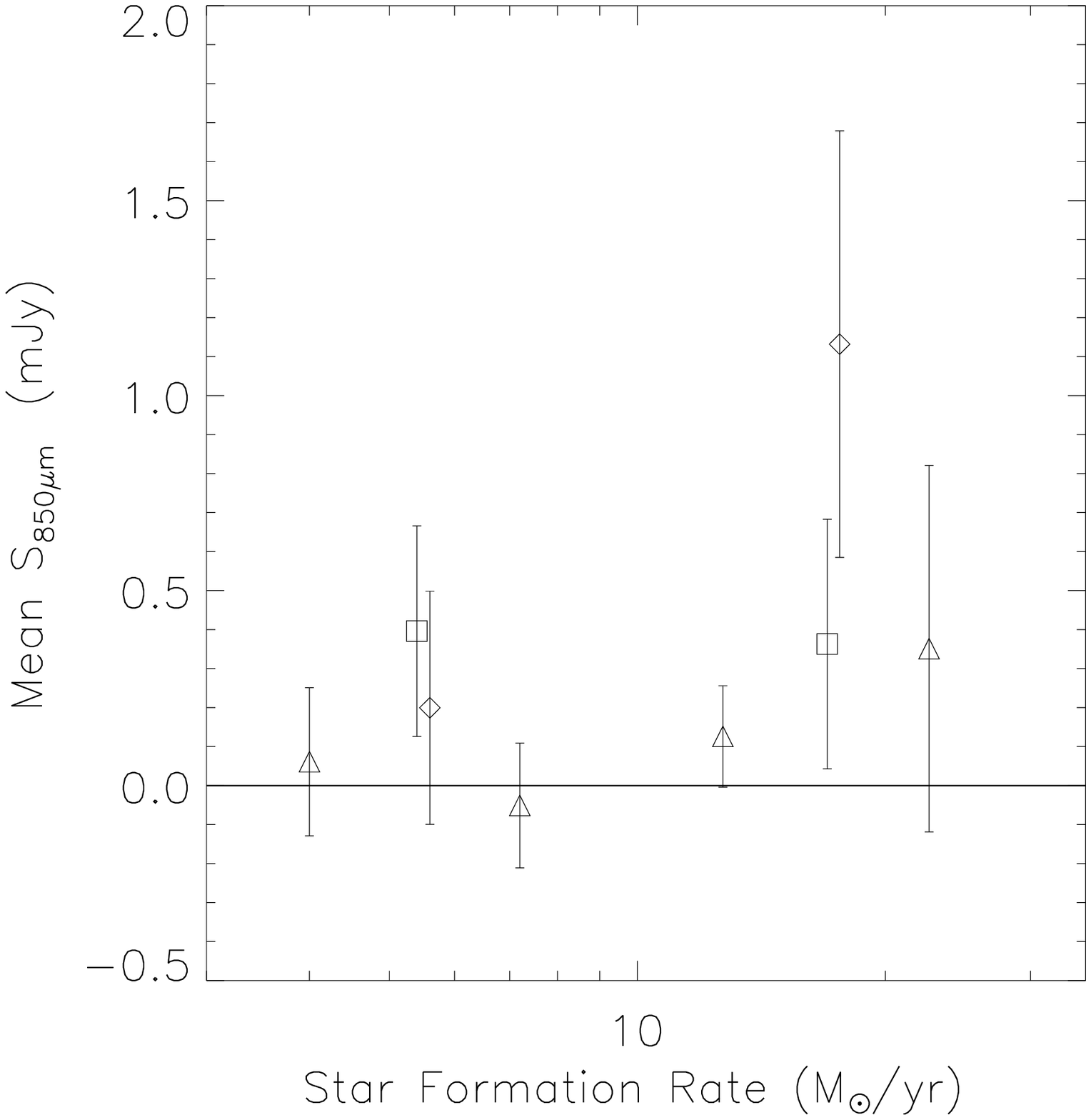}
\caption{The mean 850${\mu}$m flux as a function of $UV$-estimated star formation rate. The squares 
and diamonds correspond to the CFDF-14 and CFDF-03 samples respectively.  The triangles 
represent 
the Steidel-14 list.  Because of our small sample size we are restricted to only two bins for the CFDF 
fields 
and four for the Steidel field. The points are consistent with no detected rise in flux with star formation 
rate. }
\end{figure}

We estimate that the LBGs from the two CFDF samples could be producing at most 20\% of the 
background at 850$\mu$m, when integrated over 
all redshifts.  The Steidel LBGs, which are a deeper sample have an upper limit to their contribution of  
19\%.  However, as outlined in \citet{cha00}, 
the LBGs that are most likely do dominate the sub-mm background, the highly reddened galaxies, are 
less 
likely to be detected in these optical LBG 
surveys and therefore our sample may be biased to sub-mm faint objects.

\subsection{The SCUBA-LBG Cross-Correlation Function}

The angular cross-correlation functions between the SCUBA sources and the three LBG samples are 
presented in Figure 4.  There are some 
interesting results  from this clustering analysis.  The first is that, though we measure consistent 
clustering amplitudes for all three samples,  our strongest LBG-SCUBA 
cross-clustering signal is detected in the Steidel et al. sample 
in the 14-hour field. Recent  results from \citet{gia01}  and Foucaud et al. (in preparation) have 
shown clustering segregation with $UV$ 
luminosity for the auto-correlation function of the LBGs, such that the brighter objects are more 
strongly 
clustered. Thus one might posit that we 
should see  a stronger clustering signal in the CFDF samples. Unfortunately,  the smaller numbers in 
the brighter CFDF samples mean that even if a stronger signal was present it would be harder to 
detect.       \

A second result is that the measured amplitude of the cross-clustering  between the 14-hour SCUBA 
sources and the Steidel LBGs is larger than for the 
auto-clustering of the LBGs themselves, though certainly consistent within the error.   The strength of 
the LBG-SCUBA  angular cross-clustering is essentially a lower 
limit on the true spatial clustering as it is projected over a broad redshift range.  The actual spatial 
cross-clustering is  expected to be even higher, since the range in the redshift distribution of the 
SCUBA sources is 
 much broader than that of the LBGs.  As a further 
check we repeated the cross-correlation analysis after removing SCUBA sources with secure IDs with 
$z<$ 2 (\citet{eal00}, Webb et al, in preparation, Clements et al, in preparation)  and found the 
correlation signal  increased as would be expected for real clustering. \

By assuming a redshift distribution for the SCUBA sources (including those with $z<$ 2) we can 
estimate  
$r_{\circ}$ for the spatial LBG-SCUBA 
cross-correlation function following the procedure in (for example) \citet{efs91}.   Although the 
redshift 
distribution is still highly uncertain, the 
results of many groups are not inconsistent with a median redshift  near $z$=3 and with only a small 
fraction of objects below $z$=2  
(see the review by \citet{dunl01} and references therein).  We therefore take our general redshift  
distribution to be a Gaussian centered at $z$=3 
with 
$\sigma$=0.8.  We adopt the redshift 
distribution of \citet{gia98} for the Steidel LBGs and    ${\Omega}_m$=0.3 and 
${\Omega}_{\Lambda}$=0.7 
cosmology.  \
Given these parameters we find $r_{\circ}$=11.5 $\pm$ 3.0 $\pm$ 3.0 $h^{-1}$ 
Mpc for the  Steidel et al sample.   The first error is simply the statistical error calculated from 
${\omega}({\theta})$ and ${\delta}{\omega}({\theta})$.  
The second error is systematic and is estimated by varying the redshift distribution of the SCUBA 
sources 
from $\overline{z}$=2.5 to 
$\overline{z}$=3.5, and ${\Delta}z$ from 0.6 to 1.1.   
The redshift distribution of the  CFDF LBG samples is slightly different than that of Steidel et al.  We 
estimate the CFDF distribution as a Gaussian 
of the same standard deviation (0.24) but centered at $z=$3.2.  For the CFDF-14 amplitude of 1.1 
$\pm$  
4.4 arcsec$^{0.8}$ we find 
$r_{\circ}$=4.5 $\pm$ 7.0 $\pm$ 5.0  h$^{-1}$ Mpc and for CFDF-3 (with an amplitude of 2.3 
$\pm$3.8 
arcsec$^{0.8}$) we find $r_{\circ}$=7.5
$\pm$ 7.0 $\pm$ 5.0 $h^{-1}$ Mpc.   \

The two values of $r_o$ determined from the SCUBA-CFDF results are comparable (within their large 
uncertainties) to those measured for the 
LBG  auto-correlation function in the two fields.  In the CFDF, for ${\Omega}_m$=0.3 and 
${\Omega}_{\Lambda}$=0.7 
cosmology, Foucaud et al. (2002, in 
preparation) estimated  $r_o$ = 6.4 $\pm$ 0.3 $h^{-1}$ Mpc for the 3-hour field and $r_o$ = 5.1 
$\pm$ 
0.5 $h^{-1}$ Mpc for the 14-hour field. 
\

We consider the Steidel LBG-SCUBA cross-correlation result to be the most secure.  For Steidel 
LBGs with $\mathcal{R} < $ 25.5, \citet{gia01} 
found $r_o$ = 3.2 $\pm$ 0.7 for the auto-correlation function, for  the same cosmology which is  
smaller than our result for this field, though within the uncertainty range. One simple argument 
suggests 
that  a higher value for the amplitude of the LBG-SCUBA cross-correlation function than for
the Lyman-break galaxies themselves would  not be unexpected. The high value of the 
auto-correlation function for Lyman-break galaxies alone has been 
explained by the large values of bias expected for rare systems in the early universe \citep{kai84,gia98}.  
Indeed, the more luminous LBGs have been 
shown to be more highly clustered \citep{gia01}.  Luminous SCUBA sources are much rarer objects 
than 
Lyman-break galaxies, and so the bias 
values may  be even higher. This argument breaks down if SCUBA sources are the result of rare or 
short-lived  stochastic processes in the universe, in which case they need not  be highly-clustered at 
all, with themselves or with the LBGs.\

\citet{web02} and \citet{sco02} have measured the auto-clustering strength of SCUBA sources. 
Though 
hampered by small areas and numbers the results are consistent with strong auto-clustering, at least as 
strong 
as the auto-clustering of LBG galaxies and EROs.  \citet{web02} find an angular correlation amplitude 
of 4.4$\pm$2.9 
$\theta^{-0.8}$.  Assuming the same redshift distribution as above this may be inverted to a spatial 
amplitude of 12.8$\pm$7.0 $h^{-1}$Mpc.
 
It is tempting to draw an analogy with the
universe at low redshift, where the amplitude of the correlation function for rare clusters of galaxies is 
$\sim$18 times higher than that for
galaxies themselves, with the amplitude of the cross-correlation function between clusters and galaxies 
being midway between these
values \citep{bac88}. Circumstantial evidence in favour of this idea is the 
discovery of clusters of submillimetre sources around the
extremely rare high-redshift radio galaxies \citep{ivi00}, suggesting again that SCUBA  sources 
preferentially form in much rarer 
environments than the more Lyman-break galaxies.  However, from the results of this work we may 
only say that the cross-clustering between the SCUBA sources and LBGs is at least consistent with 
the strengths of the self-clustering in the individual populations.\

A cross-clustering signal has 
also been detected between very bright SCUBA objects and Chandra sources \citep{alm01} with an 
even 
larger amplitude than found in this work.  
X-ray bright objects, as with sub-mm bright objects are relatively rare and therefore, following the 
same 
logic as above should also be very highly 
clustered.   \

The possible angular cross-clustering signal measured in the Steidel 14-hour field is suggestive 
evidence that many of the SCUBA sources in this field are indeed at the same high-redshifts as the 
LBGs.  The smaller angular cross-clustering signal in the 3-hour field may indicate that the SCUBA 
sources in this field do not lie at the same redshifts as the LBGs and would suggest a difference in the 
redshift distribution of SCUBA sources in the 3 and 14-hour fields. \

 The positive clustering signal  should 
be remembered when attempting to 
determine identifications for SCUBA sources by positional coincidence or when observing individual 
LBGs 
with a large-beam telescope such as the 
JCMT.  Positive sub-mm flux could erroneously be associated with a near-by LBG rather than the 
object actually producing the emission which might be  undetected at optical 
wavelengths.  Indeed, this may be  the case with our own LBG 
identifications and those of other authors, in particular 
sub-mm bright LBGs which do not show unusual colours or luminosities. \

\begin{figure}
\figurenum{6(a)}
\epsscale{1.0}
\plotone{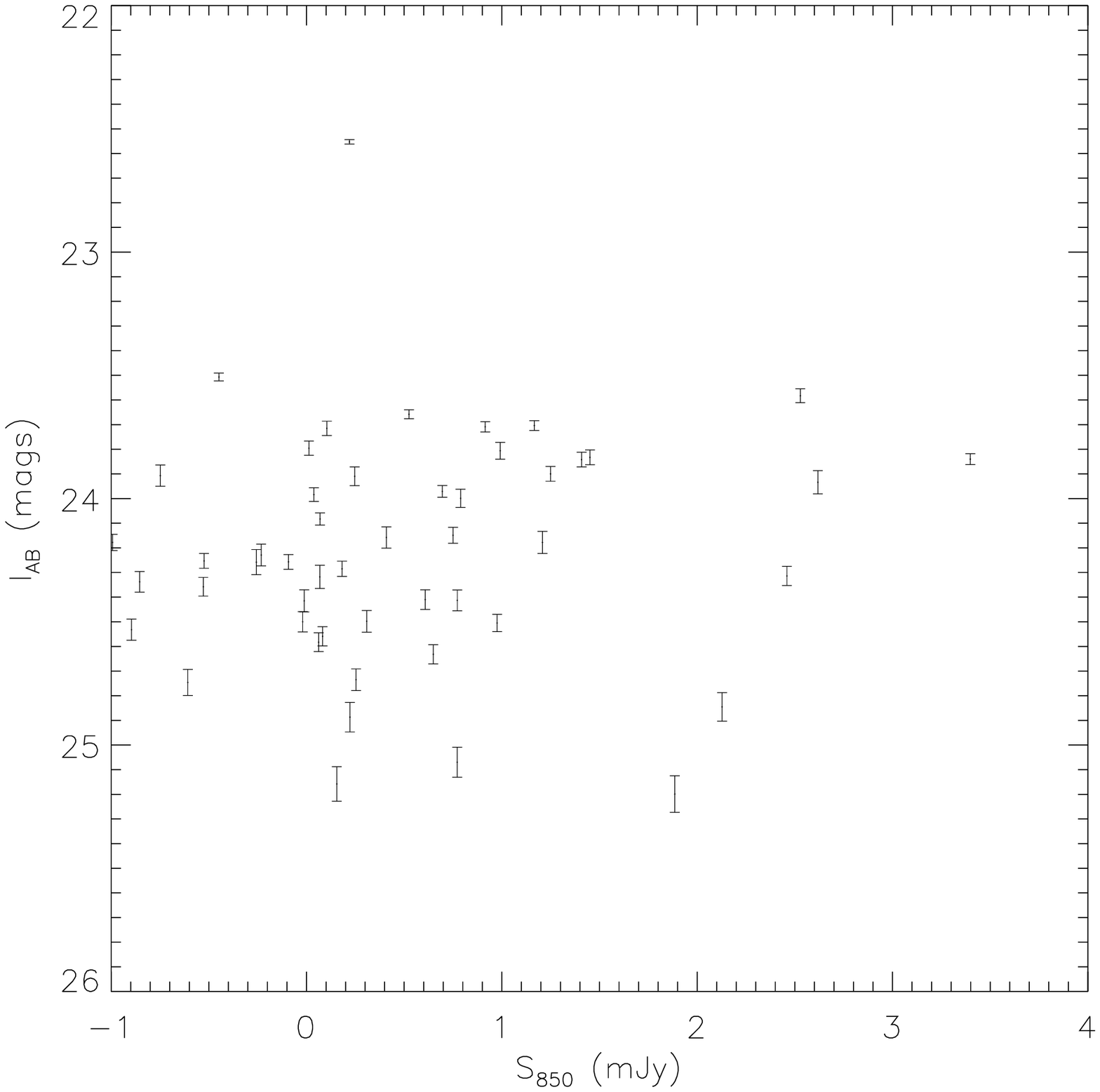}
\caption{The I-band flux of the CFDF LBGs for both the 3-hour and 14-hour fields versus their 
recovered 850 $\mu$m flux. }
\end{figure}


\begin{figure}
\figurenum{6(b)}
\epsscale{1.0}
\plotone{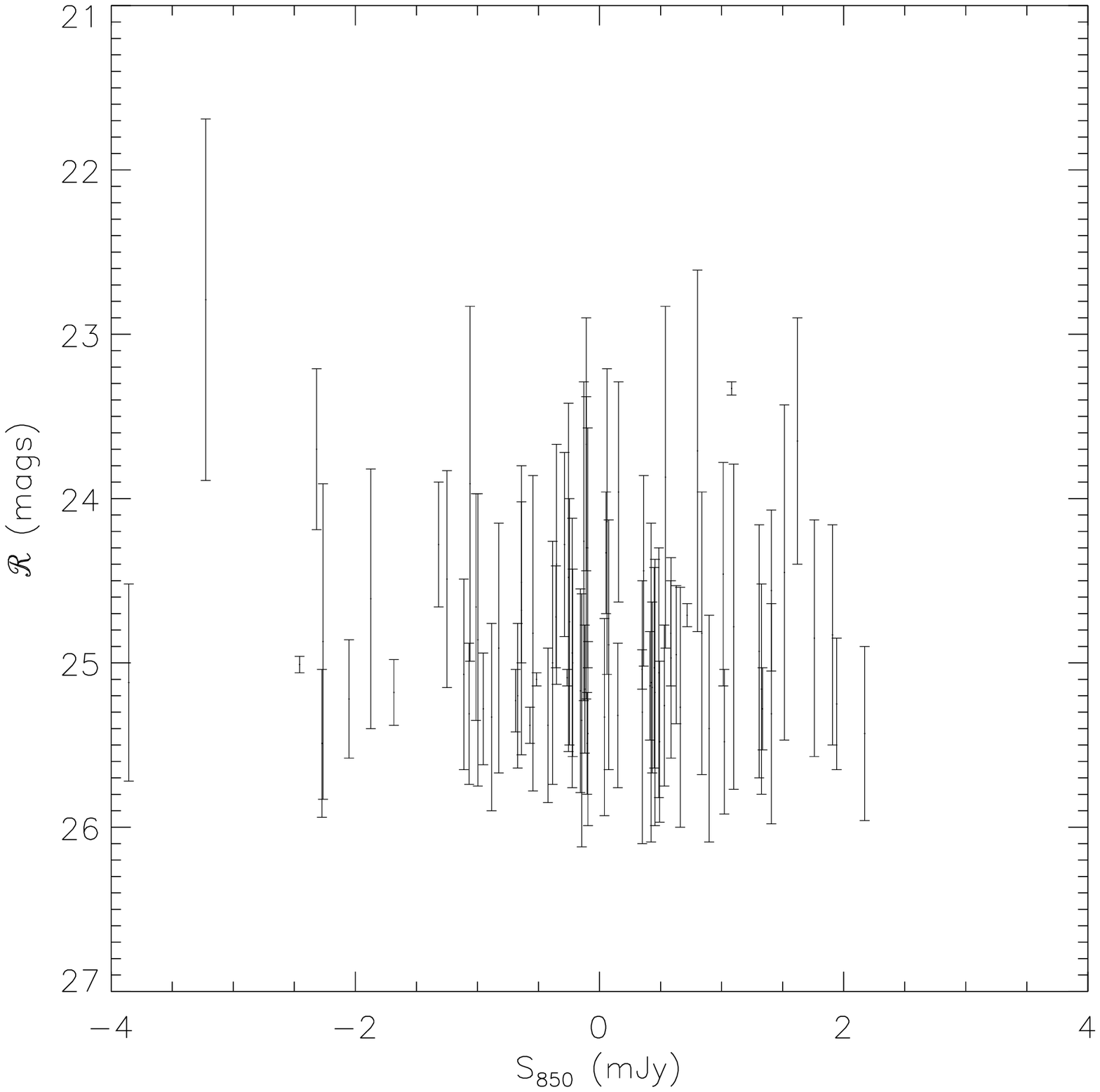}
\caption{The R-band flux of the Steidel et al 14-hour field LBGs versus their recovered 850 $\mu$m 
flux. 
} 
\end{figure}

\section{Conclusions}
We have used the 850 $\mu$m maps from the Canada-UK Deep Sub-mm Survey to study (i) the 
sub-mm 
flux and dust properties of 
Lyman-break galaxies and  (ii) the angular correlation between Lyman-break galaxies and SCUBA 
sources.  We obtain the following results: \

\

1. We marginally detect (at the 2$\sigma$ level) sub-mm flux from Lyman-break galaxies in the 
CFDF-14 
and CFDF-03 samples but we do not 
detect flux from the Steidel et al sample. The flux levels are: 0.382 $\pm$ 0.206 mJy for the 14-hour 
field and 0.414 $\pm$ 0.263 mJy for the 
3-hour 
field.  Further, we show that possibly because of LBG-SCUBA clustering SCUBA sources not 
identified 
with a LBG galaxy must be removed from 
the map before a proper analysis can be performed. \

2. Lyman-break galaxies are the best optical identification for four SCUBA sources although it is 
possible 
that some of these identifications may be 
incorrect.  There are indications that these objects may lie in a region of spatial over-density.   \  

3.  An upper limit for the dust mass of Lyman-break galaxies was calculated from their sub-mm flux 
results 
and we conclude that these masses can 
be no larger that those of near-by galaxies. \

4.  The SCUBA-LBG correlation function was measured for all three sample of Lyman-break galaxies.   
We found a high-amplitude $r_o$= 11.5 
$\pm$ 3.0 $\pm$ 
3.0  $h^{-1}$ Mpc for the Steidel et al sample, $r_o$ = 4.5 $\pm$ 7.0 $\pm$ 5.0 $h^{-1}$ Mpc for 
CFDF-14 and $r_o$ = 7.5 $\pm$ 7.0 $\pm$ 
5.0  $h^{-1}$ Mpc for CFDF-03, (where the first error is statistical and the second systematic).   

\ 

{\it Acknowledgments}  
We are grateful to the many members of the staff of the Joint Astronomy Centre who have helped us 
with 
this project.  Research by Simon Lilly is 
supported by the National Sciences and Engineering Council of Canada and by the Canadian Institute 
of 
Advanced Research. Research by Tracy 
Webb is supported by the National Sciences and Engineering Council of Canada,the Canadian 
National Research Council and the Ontario Graduate Scholarship Program.  Research by 
Stephen Eales, David Clements, Loretta Dunne and Walter Gear is supported by the Particle Physics 
and 
Astronomy Research Council.  Stephen Eales also acknowledges support from Leverhulme Trust. The 
JCMT 
is operated by the Joint Astronomy Centre on behalf of the UK Particle Physics and Astronomy 
Research 
Council, the Netherlands Organization 
for Scientific Research and the Canadian National Research Council.   We also thank Ray Carlberg for 
many helpful discussions.   \


\end{document}